\documentclass[pre,superscriptaddress,twocolumn,showpacs]{revtex4}
\usepackage{graphicx}

\begin{document}

\title{Optimized reduction of uncertainty in bursty human dynamics}
\author{Hang-Hyun Jo}
\email{hang-hyun.jo@aalto.fi}
\affiliation{BECS, Aalto University School of Science, P.O. Box 12200, Finland}
\author{Eunyoung Moon}
\affiliation{Department of Economics, University of Essex, United Kingdom}
\author{Kimmo Kaski}
\affiliation{BECS, Aalto University School of Science, P.O. Box 12200, Finland}

\begin{abstract}
Human dynamics is known to be inhomogeneous and bursty but the detailed understanding of the role of human factors in bursty dynamics is still lacking. In order to investigate their role we devise an agent-based model, where an agent in an uncertain situation tries to reduce the uncertainty by communicating with information providers while having to wait time for responses. Here the waiting time can be considered as cost. We show that the optimal choice of the waiting time under uncertainty gives rise to the bursty dynamics, characterized by the heavy-tailed distribution of optimal waiting time. We find that in all cases the efficiency for communication is relevant to the scaling behavior of the optimal waiting time distribution. On the other hand the cost turns out in some cases to be irrelevant depending on the degree of uncertainty and efficiency.
\end{abstract}

\pacs{87.23.Ge,89.65.-s,89.90.+n}
\maketitle

\section{Introduction}

Recently, significant amount of digital data on the behavior of individuals has enabled us to quantitatively explore the various patterns of human dynamics. One of the robust findings is that human dynamics is not random but correlated, such that the bursts of rapidly occuring events alternate with long inactive periods~\cite{Eckmann2004,Barabasi2005,Harder2006,Goncalves2008,Zhou2008,Radicchi2009,Karsai2011}. The bursty dynamics is typically characterized by the heavy tailed or power-law distributions of waiting time or inter-event time $\tau$ as $P(\tau)\sim\tau^{-\alpha}$ with $\alpha\approx 1$ and $0.7$ for e-mail and mobile phone call (MPC) communications~\cite{Barabasi2005,Karsai2011}, respectively. The origin of bursts in human dynamics has been debated for the past few years. The bursts can emerge from the individual selection of tasks from the task list according to the protocol ``the highest priority task first''~\cite{Barabasi2005,Vazquez2006}. The long waiting time of a task with low priority leads to the heavy tail of the waiting time distribution. This model assumes that there are at least two tasks to be compared, which is not necessarily the case in reality. On the other hand  the circadian and weekly activity patterns of humans can also result in the long periods of inactivity that contribute to the heavy tails~\cite{Malmgren2008}. In this approach the circadian and weekly patterns have been modeled by the non-homogeneous Poisson process, implying that human behavior is essentially random. While both models are based on intuitive and to some extent realistic assumptions, they are not able to consider other human related factors properly, such as how individual agents reply to the received e-mail or phone call, and what is the benefit or cost of communication.

In order to investigate the role of human factors in bursty human dynamics we devise a model, where an individual agent reacts upon the lack of information, such that an information seeking agent sends an e-mail or makes a phone call to an information provider. With this model we describe the time frame of how long people wait for responses from the information providers under uncertain (risky) situations and get insight into which human factors in the daily communication are relevant to the bursty dynamics. Here we start with an assumption that an agent prefers a fixed amount of payoff over a risky lottery of the same expected payoff, which is called \textit{risk-aversion}. A risk-averse agent can reduce uncertainty by communicating for information, which is a time-consuming process. Once an agent requests information, we consider the time to wait for responses, i.e. the waiting time, as a cost. This reflects a well-known economic perspective that time and information are considered as tradable goods~\cite{Nichols1971,Barzel1974}. Based on the trade-off between the information gain and the cost of the waiting time, the agent chooses an optimal waiting time. Note that we confine our model to risk-averse agents because the optimal waiting time for risk-pursuing agents is always zero and the waiting time does not matter for risk-neutral agents unless the information gain is unequal to the cost of time. By optimizing the waiting time, we derive a specific (positive) relationship between the risk and the optimal waiting time. Then, we show that the optimal waiting time distribution follows a power-law, which serves as theoretical support for the observed bursty dynamics in e-mail and mobile phone call communications.

Our model gives insight into the human communication dynamics, i.e. how people are concerned with the communication efficiency and the cost per unit time while deciding the communication channel as well as the time to wait for responses. Since the efficiency and the cost per unit time are heterogenous in different communication channels, the optimal channel depends on how urgent the situations are and how patient the users are. Furthermore, the efficiency and the cost per unit time also change the optimal choice of the waiting time. We find that if the cost per unit time dominates over the efficiency, people are concerned with both the efficiency and the cost per unit time under the low risk situation, while they consider only the efficiency under the high risk situation. This explains the effect of human factors on communication such that in riskier situation people tend to focus more on reducing the risk without concerning the cost, because shortening the waiting time by the efficient communication is preferred to concern the cost itself. However, if the cost per unit time is dominated by the efficiency, the result is opposite. People are concerned with only the efficiency under the low risk situation and consider both the efficiency and the cost per unit time under the high risk situation, which implies that people do not consider the cost per unit time seriously under the low risk (short waiting time) but do under the high risk (long waiting time). In conclusion the individual agents' attitude towards risk can be one of the explanations for the pattern of the waiting time in communication.

In Section~\ref{sect:model} we describe and analyze the model and explain the implications of the results. Then we make conclusions in Section~\ref{sect:conclusion}.

\section{Model}\label{sect:model}

As in~\cite{Pratt1964} we consider the risk-averse utility function, $u(x)$ of state $X$, whose degree of risk-aversion is measured by $A(x)=-u''(x)/u'(x)$. For simplicity, we set $A(x)=1$ and then consider a risk-averse agent with utility function given by
\begin{equation}
    u(x_t)=-e^{-x_t}+a\ (a\geq 0),
\end{equation}
corresponding to a time-dependent state $X_t$. At the beginning $t=0$, the state $X_0$ is uniformly distributed from $-\infty$ to $\infty$. Then, after waiting time $t>0$ for information about the state $X$, the agent reduces the uncertainty on the state as $X_t\sim N(0,\sigma^2/t^\gamma)$. That is, the state becomes more specified. The decreasing speed of uncertainty is controlled by the parameter $\gamma\geq 0$, which we call \textit{efficiency}.

As an example let us consider a job applicant who got offers from 5 firms, indexed by $i\in \{1,2,3,4,5\}$ and let $X_0=\{w_1, w_2, w_3, w_4, w_5\}$ be the set of wages of firms, which is the initial ``state'' if the applicant is uninformed. He chooses one of them and his utility depends on the wage of the firm chosen such that $u(w_i)=-e^{-w_i}+a$. Then suppose that the applicant can obtain information on the unknown wages. If the information he will get at time $t=1$ is the exact amount of $w_1$, the state becomes $X_1=\{w_2, w_3, w_4, w_5 \}$. At time $t=1$, the applicant can make a clearer decision than he could at time $t=0$ because the number of unknown wages at $t=1$ is 4. As the applicant knows the wages one-by-one, his decision gets improved, and eventually at time $t=5$, he will know the firm of the highest wage so that the best decision without uncertainty can be made.

The time the agent waits for information to reduce the uncertainty can be considered as a cost. Let $c(t)$ be the cost of time $t$. For simplicity, we assume that the cost per unit time is constant but dependent on the uncertainty level $\sigma^2$ such that $c'(t) = k\sigma^{-2/\beta}$ with positive $k$ and $\beta$. This assumption describes that given the level of the uncertainty $\sigma^2$, the cost of time is constantly increasing in time $t$ at the rate of uncertainty except for the case with $\beta=\infty$. The larger value of $\beta$ corresponds to the larger cost of the time unit, indicating that the parameter $\beta$ controls \textit{the cost per unit time}.

Now we define the agent's expected utility at state $X_t$ subtracted by the cost of time as follows:
\begin{eqnarray}
    \Pi(t)&=&E[u(x_t)]-c(t)\nonumber\\
    &=&a -\int_{-\infty}^\infty dx \sqrt{\frac{t^\gamma}{2\pi\sigma^2}} \exp \left(-x -\frac{x^2t^\gamma}{2\sigma^2}\right)-c(t)\nonumber\\
    &=&a - \exp\left(\frac{\sigma^2}{2t^\gamma}\right) -c(t).\label{eq:expectUtil}
\end{eqnarray}
The agent chooses the optimal waiting time for information. At $t=\tau$ maximizing Eq.~(\ref{eq:expectUtil}), the following condition is satisfied:
\begin{equation}
    \frac{\gamma}{2}\frac{\sigma^2}{\tau^{\gamma+1}} \exp\left(\frac{\sigma^2}{2\tau^\gamma}\right)=c'(\tau)=k\sigma^{-2/\beta}.
    \label{eq:maxCondition}
\end{equation}
From this equation and by means of Lambert function $W(x)$, satisfying $x=W(x)e^{W(x)}$, we obtain the optimal waiting time $\tau$ as a function of $\sigma$, as follows
\begin{equation}
    \tau(\sigma)=C_1 \sigma^{2/\gamma} W\left(C_2 \sigma^{2(\beta-\gamma)/[\beta(\gamma+1)]}\right)^{-1/\gamma},\label{eq:tau_exact}
\end{equation}
where $C_1\equiv [\frac{\gamma}{2(\gamma+1)}]^{1/\gamma}$ and $C_2\equiv \frac{\gamma}{2(\gamma+1)}(\frac{2k}{\gamma})^{\gamma/(\gamma+1)}$.

We first consider the special case with $\beta=\gamma$, i.e. of balancing the efficiency and the cost per unit time, leading to
\begin{equation}
    \tau(\sigma)=\tau_c \sigma^{2/\gamma}
\end{equation}
with $\tau_c\equiv C_1 W(C_2)^{-1/\gamma}$. By using the identity of $P(\tau)d\tau=P(\sigma)d\sigma$ and by assuming the distribution of $\sigma$ to be $P(\sigma)=e^{-\sigma}$, we get the optimal waiting time distribution $P(\tau)$ as
\begin{eqnarray}
    P(\tau)&=&\frac{\gamma}{2\tau_c^{\gamma/2}}\tau^{\gamma/2-1}e^{-(\tau/\tau_c)^{\gamma/2}}\\
    &\propto& \tau^{-\alpha}e^{-(\tau/\tau_c)^{\gamma/2}},
\end{eqnarray}
with the power-law exponent $\alpha\equiv 1-\gamma/2$ and the cutoff waiting time $\tau_c$. It turns out that $\alpha\leq 1$, which requires the strong cutoff such as the (stretched) exponential one as we have assumed.

Due to the technical differences among communication channels, the optimal waiting time $\tau$ can be channel-oriented. For example, one might wait longer for e-mail responses than for face-to-face responses, and e-mail responses might be faster than post responses. Thus, it is plausible that when people choose communication channels, they might expect different waiting time to different channels and choose a suitable channel to the level of uncertainty.

This channel preference can be considered in two ways: the efficiency and the cost per unit time. Firstly, the parameter $\gamma$ can be interpreted as the channel-oriented sensitivity on uncertainty. Based on the empirical observation of $\alpha\approx 1$ and $0.7$ for e-mail and MPC communications, respectively, one can argue that the corresponding values of $\gamma$ would be $\approx 0$ and $0.6$. This implies that e-mail communication rarely reduces uncertainty as e-mail users wait time for information. On the other hand, mobile phone users can reduce uncertainty more efficiently. In terms of the cost per unit time, the e-mail and the MPC can be compared as follows:
\begin{equation}
    c_{\rm e-mail}'(\tau)\approx 0 < c_{\rm MPC}'(\tau)\approx \frac{0.6}{2}\frac{\sigma^2} {\tau^{1.6}}\exp\left(\frac{\sigma^2}{2\tau^{0.6}}\right)\nonumber
\end{equation}
for $\tau>0$. Since the cost per unit time for e-mails is lower than that for MPCs, the empirical observations can be interpreted such that people are more patient to wait for the e-mail responses than for the MPC responses. Hence e-mail (MPC) is less (more) efficient and less (more) costly communication channel so that e-mail is more suitable than MPC for non-urgent situations and consequently people would prefer MPC when they need information urgently.

Since the Eq.~(\ref{eq:tau_exact}) cannot be expressed in terms of elementary functions, we use the asymptotic expansions of the Lambert function: $W(x)\approx x$ for $x\ll 1$ and $W(x)\approx \ln x$ for $x\gg 1$. Let us now consider the case with $\beta>\gamma$, where the cost per unit time is dominant over the efficiency. When $\sigma\ll 1$, we obtain the scaling relation
\begin{equation}
    \tau(\sigma)\propto \sigma^{\delta_1},\ \delta_1=\frac{2(\beta+1)}{\beta(\gamma+1)}.
\end{equation}
On the other hand, for the range of $\sigma\gg 1$, the scaling relation with logarithmic correction is obtained as
\begin{equation}
    \tau(\sigma)\propto \sigma^{\delta_2}\left[\ln (\sigma/\sigma_0)\right]^{-1/\gamma},\ \delta_2=\frac{2}{\gamma},\label{eq:log}
\end{equation}
where $\sigma_0\equiv C_2^{-\beta(\gamma+1)/[2(\beta-\gamma)]}$. Therefore, we find two scaling regimes as $\tau(\sigma)\sim \sigma^{\delta_1}$ for $\sigma<\sigma_\times$ and $\tau(\sigma)\sim \sigma^{\delta_2}$ for $\sigma>\sigma_\times$, respectively. Here $\sigma_\times$ denotes the crossover uncertainty and also defines the crossover waiting time $\tau_\times\equiv \tau(\sigma_\times)$. It indicates that the optimal waiting time distribution shows two scaling regimes with power-law exponents $\alpha_1=1-1/\delta_1$ for $\tau<\tau_\times$ and $\alpha_2=1-1/\delta_2$ for $\tau>\tau_\times$, respectively. Note that $\alpha_1<\alpha_2$ for any positive $\beta$ and $\gamma$.

Intuitively, the optimal behavior of individual agents is changing according to the risk. Under the low risk situation ($\sigma<\sigma_\times$), the power-law exponent $\alpha_1$ as a function of $\beta$ and $\gamma$ explains the case where people are concerned with the efficiency as well as the cost per unit time when the risk is not significantly high. On the other hand, under the high risk situation ($\sigma>\sigma_\times$), the power-law exponent $\alpha_2=1-\gamma/2$ turns out to be independent of $\beta$. Unlike for the low risk situations, people under the high risk consider only the efficiency, i.e. it is better to obtain information as quickly as possible due to the high cost of time. Note that the special condition of $\beta=\gamma$ yields $\delta_1=\delta_2=2/\gamma$, i.e. one scaling regime of optimal waiting time distributions.

\begin{figure}[!t]
    \begin{tabular}{c}
    \includegraphics[width=.9\columnwidth]{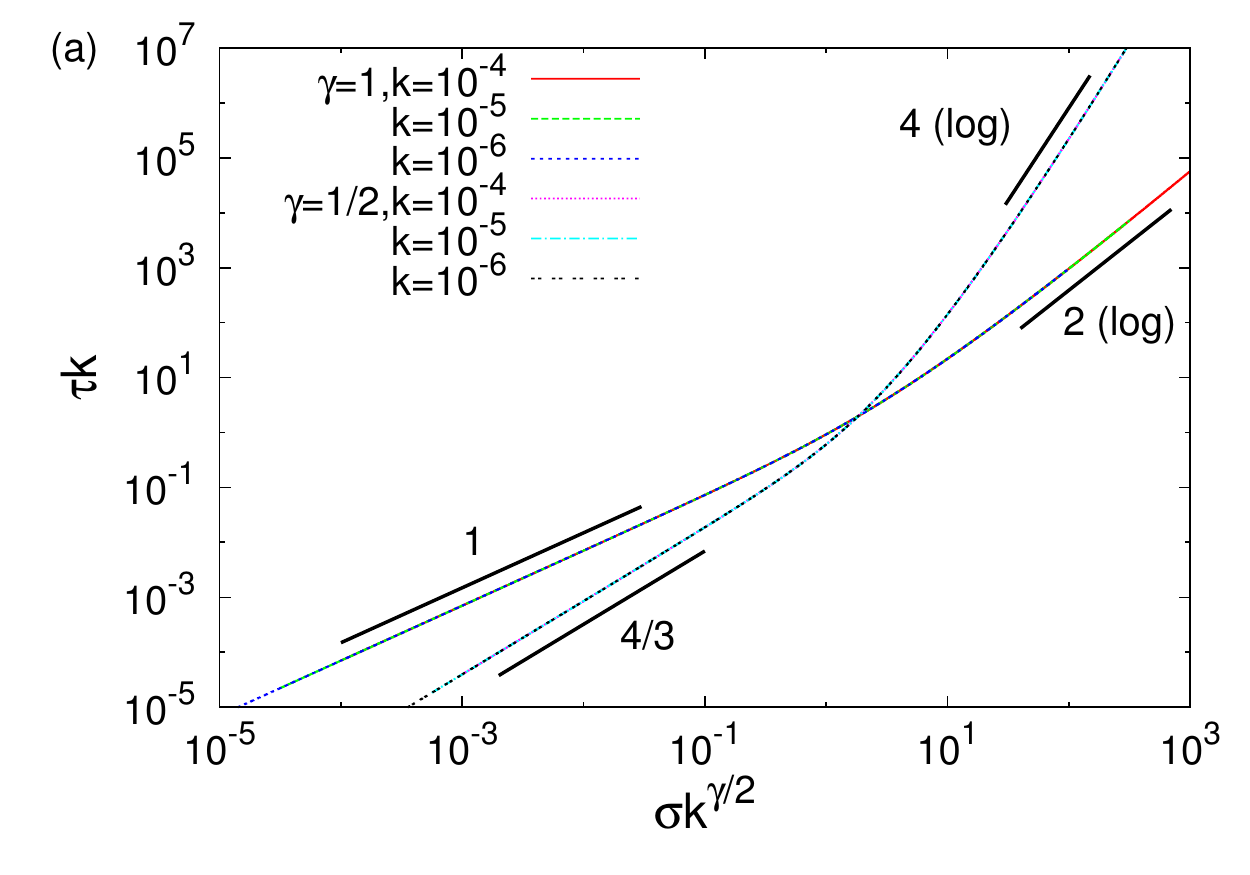}\\
    \includegraphics[width=.9\columnwidth]{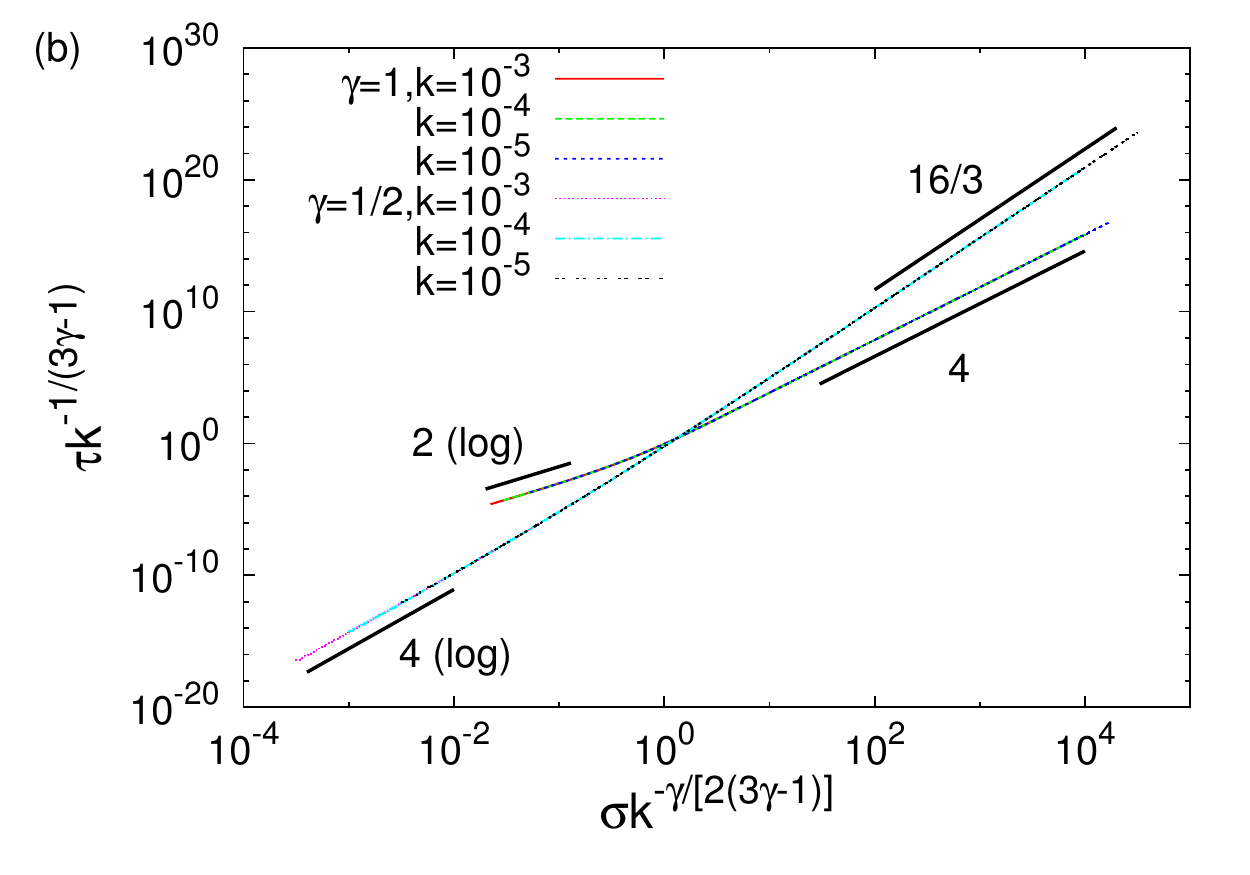}
\end{tabular}
\caption{Numerical solutions of $\tau(\sigma)$ for different values of parameters: (a) $\beta=\infty$ and (b) $\beta=1/3$. There are two scaling regimes with different power-law exponents, guided by the solid lines. The `(log)' following the exponent value denotes the fact that the power-law functions with logarithmic correction, i.e. Eq.~(\ref{eq:log}) and Eq.~(\ref{eq:largeBeta}), are used. }
\label{fig:beta}
\end{figure}

We investigate the full range of Eq.~(\ref{eq:tau_exact}) by numerically solving $\tau(\sigma)$ for various values of $k$, $\beta$, and $\gamma$. The effect of constant $k$ is systemically considered by scaling $\tau$ and $\sigma$ as $\tau k^{\beta/(\beta-\gamma)}$ and $\sigma k^{\beta\gamma/[2(\beta-\gamma)]}$, respectively. For each set of $\beta$ and $\gamma$, the solutions of $\tau(\sigma)$ for different values of $k$ collapse to the single scaling function $f$:
\begin{equation}
    \tau(\sigma)=k^{-\beta/(\beta-\gamma)}f(\sigma k^{\beta\gamma/[2(\beta-\gamma)]}),
    \label{eq:universal}
\end{equation}
where the scaling function $f$ has two scaling regimes as
\begin{equation}
    f(x)\sim\left\{\begin{tabular}{ll}
	$x^{\delta_1}$ & \textrm{if $x<x_\times$,}\\
	$x^{\delta_2}$ & \textrm{if $x>x_\times$.}
    \end{tabular}\right.
    \label{eq:fx}
\end{equation}
In the case with $\beta=\infty$, implying $c'(\tau)=k$, we obtain the numerical solutions for $\gamma=1$ and $1/2$ and for $k=10^{-4}$, $10^{-5}$, and $10^{-6}$. As shown in Fig.~\ref{fig:beta}(a), for each value of $\gamma$, the curves for different values of $k$ collapse to a single curve with two scaling regimes.

Next, we consider the case where the cost per unit time is dominated by the efficiency, i.e. $\beta<\gamma$. We find two scaling regimes of $\tau(\sigma)$ as following:
\begin{equation}
    \tau(\sigma)\propto\left\{ \begin{tabular}{ll}
    $\sigma^{\delta_1}\left[\ln (\sigma_0/\sigma)\right]^{-1/\gamma},\ \delta_1=\frac{2}{\gamma}$ & \textrm{if $\sigma<\sigma_\times$,}\\
    $\sigma^{\delta_2},\ \delta_2=\frac{2(\beta+1)}{\beta(\gamma+1)}$ & \textrm{if $\sigma>\sigma_\times$,}\\
    \end{tabular}\right.
    \label{eq:largeBeta}
\end{equation}
where $\sigma_0= C_2^{\beta(\gamma+1)/[2(\gamma-\beta)]}$. The power-law exponent for the low risk regime, $\alpha_1$, turns out to depend only on $\gamma$, while the power-law exponent for the high risk regime, $\alpha_2$, is a function of both $\beta$ and $\gamma$. In other words, when the cost per unit time is dominated by the efficiency, people are concerned with only the efficiency under the low risk situation. This is because under the low risk, implying short waiting times, the cost of waiting time is still negligible compared to efficiency. On the other hand, people under the high risk are concerned with the efficiency as well as on the cost per unit time because the cost of time is not negligible any more due to the long waiting time.

For the full range of $\tau(\sigma)$, we have numerically solved the Eq.~(\ref{eq:tau_exact}). In case with $\beta=1/3$, implying $c'(\tau)=k\sigma^{-2/3}$, we obtain the numerical solutions for $\gamma=1$ and $1/2$ and for $k=10^{-3}$, $10^{-4}$, and $10^{-5}$. As shown in Fig.~\ref{fig:beta}(b), for each value of $\gamma$, the curves for different values of $k$ collapse to a single curve of Eq.~(\ref{eq:universal}) with two scaling regimes.

\section{Conclusion}\label{sect:conclusion}

In order to investigate the role of human factors in the bursty dynamics of individuals, we have studied an agent-based model. Our model assumed an intrinsic human behavior, such as a trade-off between information gain to avoid uncertainty and the cost of time to wait for information. With the agent's optimal choice for the waiting time, the empirical waiting time distributions in daily communication process have been explained.

The results we obtained by analytical derivation show two scaling regimes for the waiting time distributions with different power-law exponents, denoted by $\alpha$. This indicates that people have different attitudes on valuing the efficiency and the cost per unit time, controlled by $\gamma$ and $\beta$, respectively. While $\alpha$ is generally expected to be a function of both $\beta$ and $\gamma$, the value of $\alpha$ turns out to be independent of $\beta$ in two cases: a) When $\beta>\gamma$ and under high risk situation, people are not concerned with the cost per unit time because it is better to obtain information as quickly as possible due to the high cost of waiting time. b) When $\beta<\gamma$ and under low risk situation, the cost of waiting time becomes negligible so that people do not have to be concerned with the cost of time unless the waiting time is significantly long.

We showed that the scaling behavior of optimal waiting time distributions reveals which human factors in the daily communication are considered as relevant to scaling. We also expect that further studies to identify intrinsic properties of social agents as well as extension of this model to interacting agents on complex networks would be helpful in figuring out the origin of bursts in human dynamics even in more details.

\acknowledgements

Financial support by Aalto University postdoctoral program (HJ), from EU's FP7 FET-Open to ICTeCollective Project No. 238597, and by the Academy of Finland, the Finnish Center of Excellence program 2006-2011, Project No. 129670 (KK) are gratefully acknowledged.

\end{document}